\documentclass[aps,prd,twocolumn,eqsecnum,showpacs,nofootinbib]{revtex4}
\usepackage{amsmath}
\usepackage[dvips]{color,graphicx}
\usepackage{amsfonts,amssymb,theorem,mathrsfs,times}

{\theorembodyfont{\upshape}

}
{\theorembodyfont{\upshape}

}
{\theorembodyfont{\upshape}

}
{\theorembodyfont{\upshape}

}
{\theorembodyfont{\upshape}

}
{\theorembodyfont{\upshape}

}
\newcommand{\dalm}{\kern1pt\vbox{\hrule height 0.9pt\hbox{\vrule width
0.9pt\hskip 2.5pt\vbox{\vskip 5.5pt}\hskip 3pt\vrule width 0.3pt}\hrule
height
0.3pt}\kern1pt}

\begin{document}

\title{ New charged black holes with conformal scalar hair }
\author{Andr{\'e}s Anabal{\'o}n$^{1,2}$}
\email{anabalon-at-aei.mpg.de}
\author{Hideki Maeda$^{2}$}
\email{hideki-at-cecs.cl}
\date{\today }

\begin{abstract}
A new class of four-dimensional, hairy, stationary solutions of the
Einstein-Maxwell-$\Lambda$ system with a conformally coupled scalar field is
obtained. The metric belongs to the Pleba{\'n}ski-Demia{\'n}ski family and
hence its static limit has the form of the charged (A)dS C-metric. It is
shown that, in the static case, a new family of hairy black holes arises.
They turn out to be cohomogeneity-two, with horizons that are neither
Einstein nor homogenous manifolds. The conical singularities in the C-metric
can be removed due to the backreaction of the scalar field providing a new
kind of regular, radiative spacetime. The scalar field carries a continuous
parameter proportional to the usual acceleration present in the C-metric. In
the zero-acceleration limit, the static solution reduces to the dyonic
Bocharova-Bronnikov-Melnikov-Bekenstein solution or the dyonic extension of
the Mart\'{\i}nez-Troncoso-Zanelli black holes, depending on the value of
the cosmological constant.
\end{abstract}

\pacs{
04.20.Jb 04.40.Nr 04.70.Bw }

AEI-2009-059, CECS-PHY-09/05

\address{
\textbf{\textit{$^{1}$}}{\small Max-Planck-Institut f\"{u}r Gravitationsphysik, Albert-Einstein-Institut, Am M\"{u}hlenberg 1, D-14476 Golm, Germany.}\\
$^{2}$Centro de Estudios Cient\'{\i}ficos (CECS), Arturo Prat 514, Valdivia, Chile
} 

\maketitle








\section{Introduction and summary}

One of the most fascinating properties of black holes is that they are
characterized only by a small set of parameters. The black hole no hair
conjecture asserts that an asymptotically flat, stationary black hole formed
from the gravitational collapse of matter fields is settled to be and
characterized only by its mass, electromagnetic charge, and angular momentum~%
\cite{rw1971}. The black hole uniqueness theorem in the asymptotically flat Einstein-Maxwell
system surely supports this conjecture, in addition to the existence of a no go result in the nonminimally coupled scalar case with quartic self interaction for a static black hole \cite{AyonBeato:2002cm}.

Among all the possible hairs, the conformal scalar hair is particularly
interesting because (i) it contains a well-known family of $U(1) $ charged
static black holes~\cite%
{Bekenstein:1996pn,Bocharova-Bronnikov-Melnikov,Bekenstein:1975ts,Virbhadra:1993st,Martinez:2002ru,Martinez:2005di}
and (ii) the asymptotically locally anti-de~Sitter (AdS) solutions in the
Einstein frame \cite{mtz2004} can be embedded in string theory~\cite%
{Papadimitriou:2007sj} and are stable against linear perturbations~\cite%
{Winstanley:2005fu}, which provide a relevant arena for the gravitational
description of superconductors~\cite{Koutsoumbas:2009pa}.

These interesting features are in contrast with the exiguous knowledge of
exact solutions of this system. The question on the existence of stationary
axisymmetric solutions was already pointed out to be of relevance in one of
the seminal papers of the subject~\cite{Bekenstein:1975ts},however, its
explicit construction has not been done until now.
The purpose of this article is to report a new exact solution in the
Einstein-Maxwell-$\Lambda$ system with a conformally coupled scalar field,
which contains all the known solutions of this system as particular limits.
A fully detailed analysis of the solution will be presented in a forthcoming paper~\cite{anabalonmaeda2}.

The exact solutions are constructed taking advantage of the following well-known fact: the traceless property of the energy-momentum tensor for a
conformally coupled scalar field implies that any spacetime with constant
Ricci scalar could support, in principle, its backreaction. Hence, the Pleba%
{\'n}ski-Demia{\'n}ski family of spacetimes~\cite{Plebanski:1976gy} (see
also~\cite{gp2006}), the most general Petrov type D spacetime in the
Einstein-Maxwell-$\Lambda$ system, provides a natural starting point.

Thus, in the next section, the most general solution in the Einstein-Maxwell-%
$\Lambda$ system with a conformally coupled scalar field within the Pleba{\'n%
}ski-Demia{\'n}ski family is constructed. The addition of a quartic
self-interaction of the scalar field is necessary to include the
cosmological constant. The subsequent section is devoted to the analysis of
the static case in order to show that all the known solutions of this system
are included within this new family as particular limits.

Our static solution, being of the form of the charged (A)dS \textit{C}-metric, is
reanalyzed in the last section to show a number of remarkable features.
First, accelerating black-hole configurations~\cite{Kinnersley:1970zw}
without conical singularities can be achieved, in contrast with the
Einstein-Maxwell-$\Lambda$ system, without implying the existence of only
two real roots in the metric functions (see~\cite{Ashtekar:1981ar}, for
instance). This is not done at the expense of changing the asymptotic
behavior of the spacetime (as opposite of the embedding of the Ernst
solution \cite{Ernst} which is asymptotic to a magnetic universe). It is
worth remarking that when the cosmological constant is present the Ernst trick to
obtain a radiative spacetime without conical singularities does not work.
The configurations that we introduce here are the first radiative solutions
that have no conical singularities. They have compact event horizons, thus
representing localized sources of matter. These configurations can be
rotating and the cosmological constant as well as a $U(1)$ gauge field can
be included, besides the scalar field, without spoiling any of these
properties.

Second, as pointed out in~\cite{Dias:2002mi}, the AdS \textit{C}-metric can be
interpreted as a single black hole in a certain range of the parameters. Our
new black hole turns out to be a cohomogeneity-two black hole whose event
horizon is neither an Einstein nor a homogenous manifold, resembling the
structure of the five-dimensional stationary black holes constructed in~\cite%
{Lu:2008js}.

Third, even in the limit where the metric has constant curvature, the
scalar field can develop a nontrivial vacuum expectation value: the
energy-momentum tensor vanishes but the scalar field is nontrivial. These
peculiar configurations have been observed to occur for Minkowski~\cite%
{AyonBeato:2005tu}, dS, and AdS spacetimes~\cite{de
Haro:2006nv,Martinez:2005di}.

The elimination of the conical singularities in the \textit{C}-metric, due to the
scalar field backreaction, is an interesting result and deserves some
comments. The conical singularities associated with the acceleration can be
neatly described as follows. The charged \textit{C}-metric can be written as~\cite%
{Kinnersley:1970zw, Ashtekar:1981ar}
\begin{align}
\mathrm{d}s^{2} =&\frac{1}{A(q-p)^{2}}\left( \frac{\mathrm{d}p^{2}}{X(p)}%
+X(p)\mathrm{d}\sigma ^{2}+\frac{\mathrm{d}q^{2}}{Y(q)}-Y(q)\mathrm{d}%
t^{2}\right) ,  \notag \\
X(p) =&1-p^{2}-2mAp^{3}-e^{2}A^{2}p^{4},\quad Y(q)=-X(q),  \label{Q}
\end{align}%
where $A$, $m$, and $e$ are acceleration, mass and charge parameters,
respectively. The manifold spanned by the coordinates $(p,\sigma )$ is
Euclidean if $X(p)\geq 0$ and compact if $X(p)$ has at least real roots.
Indeed, requiring regularity of the Killing vector field
$\partial/\partial {\sigma}$ at the degeneration surfaces one finds that either (i)
$m=0$ or (ii) $m=\pm e$ with $4Ae<-1$ or $4Ae>1$, which in turn implies
that $X(p)$ has exactly two real roots.

The situation drastically changes in the presence of the scalar field.
Slowly decaying scalar fields have nontrivial contributions to the total
mass of the spacetime~\cite{Henneaux:2004zi,Henneaux:2006hk}. Therefore, it
is in principle possible to eliminate the parameter $m$ from the metric
functions, and thus the conical singularities, still keeping the total mass
of the spacetime positive. Although this claim is not explicitly proven
below, it is supported due to the existence of solutions with four distinct
real roots even in the vanishing $m$ limit, which represent black holes free
from conical singularities.

This new family of solutions could have many applications. One of the most
interesting is, in our view, when the metric is asymptotically locally AdS
but not asymptotically static due to the acceleration horizon. The explicit
time dependence in the asymptotic region would allow the study of the elusive
nonequilibrium phenomena in the dual condensed matter system. In the
accompanying paper, we will further discuss the rotating case, its
thermodynamical aspect, and physical interpretation~\cite{anabalonmaeda2}.
Our notations follows~\cite{wald}. The conventions of curvature tensors are $%
[\nabla _\rho ,\nabla_\sigma]V^\mu ={R^\mu }_{\nu\rho\sigma}V^\nu$ and $%
R_{\mu \nu }={R^\rho }_{\mu \rho \nu }$. The metric signature is taken to be
$(-,+,+,+)$, Greek letters are spacetime indices and we set $c=1$.


\section{The stationary solution}

The Einstein-Maxwell-$\Lambda $ system with a conformally coupled scalar
field $\phi $ with quartic self-interaction can be defined by the following
set of equations:
\begin{eqnarray}
G_{\mu \nu }+\Lambda g_{\mu \nu } &=&\frac{\kappa }{4\pi }\biggl(F_{\mu \rho
}F_{\nu }^{\text{ }\rho }-\frac{1}{4}g_{\mu \nu }F_{\rho \sigma }F^{\rho
\sigma }\biggl)+\kappa T_{\mu \nu }^{(\phi )},~~~~~~  \label{eqs} \\
T_{\mu \nu }^{(\phi )} &=&\partial _{\mu }\phi \partial _{\nu }\phi -\frac{1%
}{2}g_{\mu \nu }\partial _{\rho }\phi \partial ^{\rho }\phi -\alpha g_{\mu
\nu }\phi ^{4}  \notag \\
&&+\frac{1}{6}\left( g_{\mu \nu }\kern1pt%
\vbox{\hrule height 0.9pt\hbox{\vrule width
0.9pt\hskip 2.5pt\vbox{\vskip 5.5pt}\hskip 3pt\vrule width 0.3pt}\hrule height
0.3pt}\kern1pt-\nabla _{\mu }\nabla _{\nu }+G_{\mu \nu }\right) \phi ^{2}, \\
\kern1pt%
\vbox{\hrule height 0.9pt\hbox{\vrule width
0.9pt\hskip 2.5pt\vbox{\vskip 5.5pt}\hskip 3pt\vrule width 0.3pt}\hrule height
0.3pt}\kern1pt\phi &=&\frac{1}{6}R\phi +4\alpha \phi ^{3},\quad F_{~~~~;\nu
}^{\mu \nu }=0,  \label{sc}
\end{eqnarray}%
where $\kappa :=8\pi G$, with G the Newton constant, $F_{\mu \nu }:=2\nabla
_{\lbrack \mu }A_{\nu ]}$, and $\alpha $ is a real constant. Using (\ref{sc}%
) the trace of Eq.~(\ref{eqs}) reduces to $R=4\Lambda $. Given the Pleba{%
\'{n}}ski-Demia{\'{n}}ski ansatz [the metric form~(\ref{PD}) given below],
the trace equation can be integrated to give the metric functions. Replacing
it back in the full set of field equations, we find that the most general
solution has the following form:
\begin{widetext}
\begin{eqnarray}
\mathrm{d}s^{2} &=&\frac{1}{(1-qp)^{2}}\biggl[(p^{2}+q^{2})\biggl(\frac{\mathrm{d}p^{2}}{X(p)}+\frac{\mathrm{d}q^{2}}{Y(q)}\biggl)+\frac{X(p)}{p^{2}+q^{2}}(\mathrm{d}\tau +q^{2}\mathrm{d}\sigma )^{2}-\frac{Y(q)}{p^{2}+q^{2}}(\mathrm{d}\tau -p^{2}\mathrm{d}\sigma )^{2}\biggl],  \label{PD} \\
X(p) &=& a_{0}+a_{2}p^{2}-\biggl(a_{0}+a_{4}+\frac{\Lambda }{3}\biggl)p^{4},\quad
Y(q) = a_{0}+a_{4}-a_{2}q^{2}-\left( a_{0}+\frac{\Lambda }{3}\right) q^{4}, \label{PD2}
\\
A_{\mu }\mathrm{d}x^{\mu } &=&\frac{c_{1}q+c_{2}p}{q^{2}+p^{2}}\mathrm{d}%
\tau +pq\frac{c_{2}q-c_{1}p}{q^{2}+p^{2}}\mathrm{d}\sigma ,\quad \phi =\pm \sqrt{\frac{6}{\kappa }}\frac{B(1-pq)}{C+1-pq},  \label{PD3}
\end{eqnarray}%
where the constraints on the parameters $a_{0}$, $a_{2}$, $a_{4}$, $c_{1}$, $%
c_{2}$, $B$, and $C$ depending on the values of $\alpha $ and $\Lambda $ are
summarized in Table~\ref{table0}.
\begin{table}[h]
\begin{center}
\caption{
The constraints on the parameters in the solution (\ref{PD})-(\ref{PD3}) depending on $\alpha$ and $\Lambda$. We do not consider the case with {\bf with $BC=0$}, which gives a constant scalar field. The spacetime has constant curvature for $a_{4}=0$, which we abbreviate as ``C.C.''. ``Stealth'' means that the scalar field, $\phi$, is nontrivial but its energy-momentum tensor vanishes, $T_{\mu \nu }^{(\phi)}\equiv 0$. \label{table0}}

\begin{tabular}{|c||c|c|c|c|c|}
\hline \hline
  & Constraints & Note  \\\hline
$\alpha=0$ and $\Lambda=0$& $B^2=[8\pi a_{4}-\kappa(c_{1}^2+c_{2}^2)]/(8\pi a_{4})$, $a_{4}\neq 0$, and $C=-2$  & Hairy extension of the PD spacetime \\ \hline
$\alpha=0$ and $\Lambda=0$ & $c_1=c_2=a_4=0$ and $a_0(C+2)=0$ & Stealth field on a C.C. spacetime \\ \hline
$\alpha=0$ and $\Lambda \ne 0$ & $c_{1}=c_2=a_4=0$, $a_{0}=-\Lambda(C+1)^2/[3C(C+2)]$, and $C \neq -2$ & Stealth field on a C.C. spacetime  \\ \hline
$\alpha \ne 0$ and $\Lambda = 0$ & $c_1=c_2=a_4=0$ and $B^2=-a_{0}C(C+2)\kappa/(12\alpha)$& Stealth field on a C.C. spacetime \\ \hline
$\alpha\Lambda \ne 0$& $B^2=[8\pi a_{4}-\kappa(c_{1}^2+c_{2}^2)]/(8\pi a_{4})=-\Lambda \kappa/(36 \alpha)$, $a_{4}\neq 0$, and $C=-2$  & Hairy extension of the PD spacetime \\\hline
$\alpha\Lambda \ne 0$ & $36\alpha B^2/\kappa=-\Lambda(C+1)^2-3a_{0}C(C+2)$ and $c_{1}=c_2=a_4=0$ & Stealth field on a C.C. spacetime \\
\hline \hline
\end{tabular}
\end{center}
\end{table}
\end{widetext}

The most relevant conclusion following from Table~\ref{table0} is that the
spacetime has nontrivial rotation. Indeed, for $B=0$, the scalar field
vanishes and the metric corresponds to the usual Pleba{\'{n}}ski-Demia{\'{n}}%
ski family of solutions with $8\pi a_{4}=\kappa (c_{1}^{2}+c_{2}^{2})$ and
vanishing mass and NUT (Newman-Unti-Tamburino) parameters. Thus, the metric contains the accelerated
version of the zero-mass Kerr-Newman spacetime. This fact makes us confident
that the angular momentum is not pure gauge in the above metric with $BC\neq
0$.

The family of solutions supporting a nontrivial scalar field, $BC \ne 0$,
has two branches and the parameters of the metric are accordingly related in
a different way. The first branch is when $a_{4}=c_{1}=c_{2}=0$ then the
metric has constant curvature, $R_{~~~\lambda \rho }^{\mu \nu
}=(\Lambda/3)\left( \delta _{\lambda }^{\mu }\delta _{\rho }^{\nu
}-\delta_{\rho }^{\mu }\delta _{\lambda }^{\nu }\right)$, and the parameters
are related through the single relation $36\alpha
B^2/\kappa=-\Lambda(C+1)^2-3a_{0}C(C+2)$. It follows that the scalar field
carries an integration constant and that it is a stealth field~\cite%
{AyonBeato:2005tu, de Haro:2006nv,Martinez:2005di}, namely a nontrivial
scalar field giving $T_{\mu \nu }^{(\phi)}\equiv 0$.

For $a_{4}\neq 0$, the metric is no longer of constant curvature. In the
case of $\Lambda =0=\alpha $, the above configuration is a solution with a
nonconstant scalar field if and only if $C=-2$ and
\begin{equation}
B^{2}=\frac{8\pi a_{4}-\kappa (c_{1}^{2}+c_{2}^{2})}{8\pi a_{4}}.
\end{equation}%
The above relation entails the main difference from the Pleba{\'{n}}ski-Demia%
{\'{n}}ski family with vanishing scalar field. As we remarked before, $B=0$
results in $8\pi a_{4}=\kappa (c_{1}^{2}+c_{2}^{2})$ and the parameter $a_{4}
$ in the metric functions must be strictly positive. The scalar field
relaxes this condition allowing a negative $a_{4}$. As we discuss in the
next section, in the nonrotating case, this implies the existence of a new
family of black holes that do not exist when the scalar field vanishes. For $%
\alpha \Lambda <0$, the above relation becomes
\begin{equation}
B^{2}=\frac{8\pi a_{4}-\kappa (c_{1}^{2}+c_{2}^{2})}{8\pi a_{4}}=-\frac{%
\kappa \Lambda }{36\alpha }.  \label{const}
\end{equation}%
Note that the value of $B$ is not arbitrary but fixed as $B^{2}=1$ when the
stealth configuration is obtained taking the limit $a_{4}\rightarrow 0$ from the nontrivial solution with $c_{1}=c_{2}=0$ and $%
a_{4}\neq 0$.

\section{Recovering the known solutions}

In this section, we show that the nontrivial solution, namely $C=-2$ and
the relation given in (\ref{const}), reduces to the known solutions as
limiting cases. First, we consider the static limit of our stationary
solution (\ref{PD})-(\ref{PD3}): its static limit is achieved after the
coordinate transformations $p\to p/n$, $q\to n/q$, $\sigma\to \sigma/n$, and
$\tau\to \tau/n$ together with the redefinitions of the parameters such that
$a_{2}\to n^2a_{2}$, $a_{4}\to n^4a_{4}$, $c_1\to n^2c_1$, and $c_2\to
n^2c_2 $ and the limit $n \to \infty$. The further coordinate
transformations $p\to \beta p-a_{3}/(4a_{4})$, $q\to q-a_{3}/(4a_{4})$, and $%
\sigma\to \sigma/\beta $ and redefinitions $a_{0}\to
\beta^2a_{0}-(16a_{2}a_{4}-a_{3}^2)a_{3}^2/(256a_{4}^3)$ and $a_{2}\to
a_{2}+3a_{3}^2/(8a_{4})$ bring the solution to the form of [modulo a gauge
transformation of the $U(1)$ field]
\begin{widetext}
\begin{eqnarray}
\mathrm{d}s^{2} &=&\frac{1}{(q-\beta p)^{2}}\left[ \frac{\mathrm{d}q^{2}}{%
Y(q)}+\frac{\mathrm{d}p^{2}}{X(p)}-Y(q)\mathrm{d}\tau ^{2}+X(p)\mathrm{d}\sigma ^{2}\right],\qquad \phi =\sqrt{\frac{6}{\kappa }}\frac{B(\beta p-q)}{\beta p+q-a_{3}/(2a_{4})},\label{P3}\\
X(p) &=&a_{0}+\frac{a_{1}}{\beta }p+a_{2}p^{2}+\beta a_{3}p^{3}-\beta
^{2}a_{4}p^{4}, \qquad
Y(q) =-\beta ^{2}a_{0}-a_{1}q-a_{2}q^{2}-a_{3}q^{3}+a_{4}q^{4}-\frac{%
\Lambda }{3},  \label{P42} \\
a_{1} &=&-\frac{a_{3}\left( 4a_{2}a_{4}+a_{3}^{2}\right) }{8a_{4}^{2}}, \qquad
A_{\mu }\mathrm{d}x^{\mu } =c_{1}q\mathrm{d}\tau +c_{2}p\mathrm{d}\sigma ,\label{P5}
\end{eqnarray}
\end{widetext}
where $a_{4}\neq 0$ is assumed and new parameters $a_{1}$, $a_{3}$,
and $\beta $ were introduced. They allow considering the zero-acceleration
limit, $\beta \rightarrow 0$. It can be noted that if no coordinate
transformations is done after taking the static limit, $n\rightarrow \infty $%
, the configuration (\ref{P3})-(\ref{P5}) would have been in the same form
but with $\beta =1$ and $a_{1}=a_{3}=0$. It follows that we can set $%
|a_{2}|=1$ or $|a_{4}|=1$ if $a_{2}a_{4}\neq 0$ using a remaining degree of
freedom $p\rightarrow dp$, $q\rightarrow dq$, $\tau \rightarrow d\tau $, $%
\sigma \rightarrow d\sigma $, $c_{1}\rightarrow c_{1}/d^{2}$, and $%
c_{2}\rightarrow c_{2}/d^{2}$ with a constant $d$. Hence, for $\beta \neq 0$%
, there are five independent parameters.

Let us consider now the zero-acceleration limit $\beta \rightarrow 0$ of the
static solution (\ref{P3})-(\ref{P5}). This makes sense only in the case of $%
a_1=0$, which requires $a_{3}=0$ or $a_{4}=-a_{3}^{2}/(4a_{2})$ with $a_2
\ne 0$. In the case of $a_{3}=0$, the limit implies a constant scalar field.
Note that in the previously known solutions of this system the scalar field
does not carry any continuous parameter that allows driving it to a
nonzero constant value. In the case where $a_{4}=-a_{3}^{2}/4a_{2}$, by the
coordinate transformation $r:=1/q$ and the rescaling of the coordinates $%
\tau \to \tau/\sqrt{|a_2|}$, $r \to \sqrt{|a_2|}r$, $p \to \sqrt{|a_0|}p/%
\sqrt{|a_2|}$, and $\sigma \to \sigma/(\sqrt{|a_0||a_2|})$ together with the
redefinition of the parameters such as $e:=c_1/|a_2|$, $g:=c_2/|a_2|$, $%
MG:=-a_3/(2a_2|a_2|^{1/2})$, and $k:=-\mbox{sign}(a_{2})$, the limit
provides the dyonic extension of the black hole obtained in~\cite%
{Martinez:2002ru};

\begin{align}
\mathrm{d}s^{2}=& -f(r)\mathrm{d}\tau ^{2}+\frac{\mathrm{d}r^{2}}{f(r)}+r^{2}%
\biggl(\frac{\mathrm{d}p^{2}}{\mbox{sign}(a_{0})-kp^{2}} \\
& +(\mbox{sign}(a_{0})-kp^{2})\mathrm{d}\sigma ^{2}\biggl), \\
A_{\mu }\mathrm{d}x^{\mu }& =\frac{e}{r}\mathrm{d}\tau +gp\mathrm{d}\sigma
,\qquad k\frac{e^{2}+g^{2}}{M^{2}}=G+\frac{2\pi \Lambda }{9\alpha }G^{2}, \\
f(r)=& k\biggl(1-\frac{MG}{r}\biggl)^{2}-\frac{\Lambda }{3}r^{2}, \\
\phi & =\pm \sqrt{\frac{3}{4\pi }}\frac{\sqrt{M^{2}G-k(e^{2}+g^{2})}}{r-GM},
\end{align}%
where $\alpha \Lambda <0$ is required for the scalar field to be real and $k$
represents the curvature of the two-dimensional section of $(p,\sigma )$.
Although the limit is not well-defined for $a_{2}=0$ corresponding to $k=0$,
the above solution is valid even for $k=0$, in which $e=g=0$ is required.
Thus, all the known solutions of the relevant system are contained within
the static family (\ref{P3})-(\ref{P5}).


\section{New cohomogeneity-two black holes}

Now let us see the important consequence of the scalar hair in the static
case (\ref{P3})--(\ref{P5}), where we can set $\beta=1$ and hence ${a}_1={a}%
_3=0$ without loss of generality. There are then two different families of
solutions, depending on the sign of $a_4$. For $a_4>0$, the geometry is the
same as in the extremal case of the $U(1)$ charged (A)dS \textit{C}-metric; the
relation of this case with a conformally coupled scalar field is analyzed in~%
\cite{Charmousis:2009cm}. Interestingly, the case with negative $a_4(=:-b^2)$
is possible in the presence of the scalar hair. This case does not occur
within the pure Einstein-Maxwell-$\Lambda$ system. In what follows we focus
on this case.

The $(p,\sigma)$ submanifold is Euclidean and compact if and only if $X(p)$
has four real roots. In terms of these roots the metric functions are
\begin{eqnarray}
X(p)&=&b^{2}\left( p^{2}-\xi _{1}^{2}\right) \left( p^{2}-\xi
_{2}^{2}\right), \\
Y(q)&=&-b^{2}\left( q^{2}-\xi _{1}^{2}\right) \left(
q^{2}-\xi_{2}^{2}\right) -\frac{\Lambda }{3},  \label{Y}
\end{eqnarray}
where we set $0<\xi _{1}<\xi _{2}$ without loss of generality. It follows
that the required signature and compactness is obtained if $-\xi _{1}\leq
p\leq \xi _{1}$. From the expansion of the metric around the degeneration
surfaces of the angular Killing vector $\partial/\partial\sigma$, it follows
that the spacetime is free from conical singularities, as can be seen from
the relation $|X^{\prime}(\xi_1)|=|X^{\prime}(-\xi_1)|=2b^2\xi_1(\xi_2^2-%
\xi_1^2)$, where a prime denotes the derivative. Conformal infinity is
located at $p=q$ and there are curvature singularities at $q=\pm\infty$, so
the domain of the coordinate $q$ is $p<q<\infty$.

When the cosmological constant vanishes, there is an event horizon at $%
q=\xi_{2}$ and an acceleration horizon at $q=\xi_{1}$. For $\Lambda \neq 0$,
the roots of $Y(q)=0$, $q_{1(+)}$, $q_{1(-)}$, $q_{2(+)}$, and $q_{2(-)}$,
are given by
\begin{align}
q_{\varepsilon (\pm)}:=\pm\frac{1}{\sqrt{2}}\biggl[\xi_1^2+\xi_2^2-%
\varepsilon \sqrt{(\xi_1^2-\xi_2^2)^2-\frac{4\Lambda}{3b^2}}\biggl]^{1/2},
\end{align}
where $\varepsilon=\pm 1$. $q=q_{1(+)}$ and $q=q_{2(+)}$ correspond to the
acceleration horizon and the event horizon, respectively.

Let us count the number of the real roots of $Y(q)=0$. When $%
\Lambda>3b^2(\xi_1^2-\xi_2^2)^2/4$, there is no root of $Y(q)=0$ and the
Killing vector $\partial/\partial\tau$ becomes spacelike everywhere. There
are two roots for $\Lambda=3b^2(\xi_1^2-\xi_2^2)^2/4$; here the event and
acceleration horizon coalesce. In the case of $\Lambda<3b^2(\xi_1^2-%
\xi_2^2)^2/4$, there are four, three, and two roots for $\Lambda>-3b^2\xi
_{1}^2\xi _{2}^2$, $\Lambda=-3b^2\xi _{1}^2\xi _{2}^2$, and $%
\Lambda<-3b^2\xi _{1}^2\xi _{2}^2$, respectively.

In the case of the positive or vanishing cosmological constant, the
spacetime is not static near the conformal infinity. The situation is quite
different for the negative cosmological constant. The acceleration horizon
exists only for $-3b^2\xi _{1}^2\xi _{2}^2\le \Lambda<0$ with equality
holding for the case with the extremal horizon. For $\Lambda<-3b^2\xi_{1}^2%
\xi _{2}^2$, in contrast, there is no acceleration horizon and the spacetime
is static near the conformal infinity. When the asymptotic region is static
the interpretation of the \textit{C}-metric changes and it corresponds to the
geometry of a single black hole~\cite{Dias:2002mi}. Thus, these cases
represent new asymptotically locally AdS black holes without conical
singularities.

It should be noted that, when the cosmological constant is negative, the
coordinate rank $q>p$ implies the existence of constant $p$ slices which do
not intersect the two acceleration horizons. Whenever the
acceleration horizons exist and the cosmological constant is nonpositive,
these horizons reach infinity. When the cosmological constant is positive,
the acceleration horizon is replaced by a compact cosmological horizon for
the allowed values of $\Lambda$ discussed before.

The spacetime is regular everywhere outside the event horizon. Now let us
consider the behavior of the conformal scalar field, which is given by $\phi=\sqrt{6/\kappa}B(p-q)/(p+q)$. The scalar field diverges on the surface $p+q=0$. This surface is outside
the cosmological horizon for $\Lambda > 0$. Depending on the value of $p$,
it is outside or on the acceleration horizon for $\Lambda = 0$ and outside,
on, or inside the acceleration horizon for $-3b^2\xi _{1}^2\xi _{2}^2<
\Lambda<0$. For $\Lambda=-3b^2\xi _{1}^2\xi _{2}^2$, it is completely inside
of an extremal horizon. Note that in the spherically symmetric
Bocharova-Bronnikov-Melnikov-Bekenstein black-hole, that surface is located
precisely on the event horizon~\cite%
{Bocharova-Bronnikov-Melnikov,Bekenstein:1975ts}. The scalar field reduces
to zero at the conformal infinity and is regular on the event horizon.

The horizon metric with constant $\tau $ is given by
\begin{equation}
\mathrm{d}s_{\mathrm{H}}^{2}=\frac{1}{(q_{\mathrm{H}}-p)^{2}}\left( \frac{%
\mathrm{d}p^{2}}{X(p)}+X(p)\mathrm{d}\sigma ^{2}\right) ,
\end{equation}%
where $q_{\mathrm{H}}$ is the value of $q$ at the event horizon. Note that
the horizon manifold $M_{H}$ is neither Einstein nor homogeneous. The
topology of this event horizon is defined by its Euler characteristic $\chi$%
. The lack of conical singularities implies that $\sigma \in \left[ 0,2\pi
/\{b^{2}\xi _{1}(\xi _{2}^{2}-\xi_{1}^{2})\}\right] $, from which it follows
\begin{equation}
\chi :=\frac{1}{4\pi }\int_{M_{H}}{}^{(2)}R\sqrt{g}dpd\sigma =2,
\end{equation}
where ${}^{(2)}R$ is the two-dimensional Ricci scalar of $M_{H}$.
Therefore, the horizon is diffeomorphic to a two-sphere. The metric is
asymptotically locally (A)dS in the sense that $R_{~~~\lambda \rho }^{\mu \nu }|_{p=q}=(\Lambda/3)( \delta
_{\lambda }^{\mu }\delta _{\rho }^{\nu }-\delta_{\rho }^{\mu }\delta
_{\lambda }^{\nu })$.

As a final remark we would like to stress that the parameters given in the
metric have not been labeled as mass, electric or magnetic charge because
these quantities are meaningful only when they are defined as surface
integrals. We make a more extended analysis of these issues as well as the
thermodynamical properties of these spacetimes in the forthcoming work~\cite%
{anabalonmaeda2}.

\acknowledgments
The authors would like to thank D. Astefanesei, E. Ay\'{o}n-Beato, J. Bi\v{c}\'{a}k, F.
Ferrari, S. Hartnoll, G. Lavrelashvili, C. Mart\'{\i}nez, J. Oliva, M.
Rodriguez, S. Tomizawa, R. Troncoso, J.~Podolsk{\'y}, O. Varela,
E.~Winstanley, and J. Zanelli for discussions and useful comments. A.A. wishes
to thank the precious support of Cecilia Gumucio while writing this
manuscript. This work was partially funded by the following Fondecyt grants:
3080024 (AA), and 1071125 (HM). This work was partially supported by the
Alexander von Humboldt foundation. The Centro de Estudios Cient\'{\i}ficos
(CECS) is funded by the Chilean Government through the Millennium Science
Initiative and the Centers of Excellence Base Financing Program of Conicyt.
CECS is also supported by a group of private companies which at present
includes Antofagasta Minerals, Arauco, Empresas CMPC, Indura, Naviera
Ultragas, and Telef\'{o}nica del Sur. CIN is funded by Conicyt and the
Gobierno Regional de Los R\'{\i}os.

~\newline
\emph{Note added:} At the final stage of this work, we were informed that
another group also obtained the static solution (\ref{P3})-(\ref{P5})~\cite%
{Charmousis:2009cm}. In this group's analysis, only the case where $a_4$ is
negative [corresponding to the $e^2$ term of (\ref{Q}) that is positive] is
studied. In fact, this $e^2$ term is not the square of the electric charge
and can be negative, which is the case we analyzed in the last section.


{%

}

\end{document}